\documentstyle[prl,aps,epsfig,twocolumn]{revtex}

\begin{document}
\draft

\title{High efficiency photon counting using stopped light}

\author{A. Imamo\=glu$^{1,2}$}

\address{$^1$ Department of Electrical and Computer Engineering,
University of California, Santa Barbara, CA 93106}

\address{$^2$ Department of Physics,
University of California, Santa Barbara, CA 93106}

\date{\today}


\maketitle

\begin{abstract}
Single-photon detection and photon counting play a central role in
a large number of quantum communication and computation protocols.
While the efficiency of state-of-the-art photo-detectors is well
below the desired limits, quantum state measurements in trapped
ions can be carried out with efficiencies approaching $100 \%$.
Here, we propose a method that can in principle achieve ideal
photon counting, by combining the techniques of photonic quantum
memory and ion-trap fluorescence detection: after mapping the
quantum state of a propagating light pulse onto metastable
collective excitations of a trapped cold atomic gas, it is
possible to monitor the resonance fluorescence induced by  an
additional laser field that only couples to the metastable excited
state. Even with a photon collection/detection efficiency as low
as $10 \%$, it is possible to achieve photon counting with
efficiency approaching $100 \%$.
\end{abstract}

\draft

\vskip2pc \narrowtext

Quantum information science has emerged as a truly
interdisciplinary field involving the contributions of physicists,
engineers, and computer scientists \cite{NC}. Despite the
impressive progress in theory within the last 8 years,
experimental implementation of quantum information processing in
physical systems remains as a major challenge. A significant
fraction of key experiments in the field, such as Bell's
inequality violations \cite{aspect,aspect2}, quantum key
distribution \cite{QKD} and quantum teleportation \cite{QT}, have
been carried out using photons and linear optical elements such as
polarizers and beam splitters. Even in these earlier experiments,
efficiency of single-photon detection had been a limiting factor.
Recently, Knill, Laflamme, and Milburn have shown theoretically
that efficient linear optics quantum computation (LOQC) can be
implemented using on-demand single-photon pulses and
high-efficiency photon-counters \cite{KLM01}. The required
photon-counting efficiency for LOQC is estimated to exceed $99
\%$: this is beyond the capability of state-of-the-art devices,
which can only provide single-photon detection (and not photon
counting) efficiencies below $90 \%$.

In this Letter, we propose a scheme to carry out photon-counting
with efficiency potentially exceeding $99\%$. The proposed method
combines the techniques of ion-trap quantum-state measurements
\cite{ion-trap} with that of quantum memory for photons
\cite{FL00} based on electromagnetically induced transparency
\cite{EIT}. The first step in this proposal is the mapping of the
quantum state of the propagating light pulse onto collective
excitations of an atomic gas, as described below. The number of
collective excitations are then measured efficiently using the
state-selective fluorescence measurements developed for trapped
ions. Both steps in the process can in principle be carried out
with efficiency exceeding $99 \%$, limited only by the dephasing
time of the collective atomic excitations.

Arguably, the only physical system where a quantum state
measurement with efficiency approaching $100 \%$ has been
demonstrated, is trapped ions \cite{ion-trap}. The goal of these
measurements is to determine whether an ion is in its ground state
($|g \rangle$) or in a (pre-selected) hyperfine-split metastable
excited state ($|m \rangle$). The measurement proceeds by coupling
state $|m \rangle$ to an excited fluorescing state $|f \rangle$
(that decays by spontaneous emission back to state $|m \rangle$)
using a resonant laser field. If no photon is detected, the state
of the ion collapses onto $|g \rangle$. Similarly, observation of
scattered photons (i.e. resonance fluorescence) projects the state
of the ion to $|m \rangle$. By choosing the states $|m \rangle$
and $|f \rangle$ to be the $|F=2, m_F=2 \rangle$ and $|F=3, m_F=3
\rangle$ states of the ground and excited levels respectively, one
can ensure that the decay of state $|f \rangle$ will only populate
state $|m \rangle$ and that many photons can be scattered.

Hyperfine-split states of the ground electronic level of an ion or
an atom constitute an ideal quantum bit (qubit) \cite{NC}, due to
the ultra-long decoherence time of these states. In the case of an
ensemble of atoms, these long decoherence times have also been
utilized in experiments demonstrating electromagnetically induced
transparency (EIT) \cite{hau99}. The essence of EIT is the
creation of coherence between two long lived atomic states by two
laser fields in an optically thick medium \cite{EIT}: the
absorption experienced by the optical field that couples the
ground atomic state $|g\rangle$ to a common excited state
$|e\rangle$ (Fig.1), is eliminated by the non-perturbative
coupling laser field that is applied at the $|m\rangle-|e\rangle$
transition. Elimination of resonant absorption in EIT is due to a
quantum interference effect that leads to a cancellation of
optical excitation probability amplitudes only when the probe and
the coupling fields satisfy exact Raman resonance. The narrow
transparency window for the probe field is accompanied by a very
steep variation of the refractive index with frequency, leading to
ultra-slow group velocity of probe pulses travelling in EIT medium
\cite{EIT}.

A probe laser pulse may enter an EIT medium without experiencing
any loss or reflection, provided that its bandwidth $\Delta
\nu_{probe}$ is much smaller than the width of the transparency
window $\Delta \nu$ \cite{EIT}. Under these conditions, the
incident probe pulse excites coupled collective excitations of
atoms and light inside the medium, which propagate with an
ultra-slow group velocity \cite{EIT-dispersive}. The quantum state
inside the medium that corresponds to an incoming probe pulse with
exactly n photons is \cite{FL00,FL02}
\begin{eqnarray}
| \Phi_n \rangle & = & \frac{1}{\sqrt{n!}} \,
[\frac{\Omega_c(t)}{\sqrt{\Omega_c^2(t) + g^2 N}} \,
\hat{a}^{\dagger}(z,t) \nonumber \\
 &-& \frac{g
\sqrt{N}}{\sqrt{\Omega_c^2(t) + g^2 N}} \,
\hat{\sigma}_{mg}(z,t)]^n |G\rangle \otimes |0\rangle \;,
\label{eq1}
\end{eqnarray}
where $|G \rangle = |g\rangle_1 |g\rangle_2 .....|g \rangle_N$ is
the ensemble atomic state where all $N$ atoms are in the
ground-state and $|0 \rangle$ is the vacuum state of the probe
field mode. $\Omega_c(t)$ and $g$ denote the Rabi frequency of the
coupling field and the single-atom-field interaction strength of
the probe transition, respectively. The operator
$\hat{a}^{\dagger}(z,t)$ creates a travelling probe photon;
$\hat{\sigma}_{mg}(z,t)$ is the collective atomic raising
operator. Extension of these atom-field states to arbitrary
incoming field states is straightforward \cite{FL02}. The
corresponding excitations are referred to as dark-state
polaritons, since atomic excitations take place between
(spontaneous-emission-free) states $|g\rangle$ and $|m\rangle$.

It has been first noted by Fleischauer and Lukin \cite{FL00} that
it is possible to make the dark-state polaritons purely atom-like
by adiabatically turning the coupling field off. When this
transformation is implemented, we end up with a purely atomic
collective excitation; in this limit, the group velocity of the
dark-state polariton is zero. It has also been shown in
Ref.\cite{FL00} that by turning the coupling laser back on
adiabatically, it is possible to reverse the quantum-state
transfer and re-generate the probe field pulse in the original
quantum state. This proposal has been realized experimentally
using an ultra-cold atomic sample in Ref.~\cite{hau01} , and warm
Rb atoms in Ref.~\cite{walsworth}.

The first step of the photon counting procedure that we propose is
based on this transfer between the quantum states of the probe
pulse and the metastable collective atomic excitations. After the
state transfer is completed by adiabatically turning the coupling
laser Rabi frequency $\Omega_c(t)$ off, we have a $1-1$ mapping
between the incoming photons and the collective atomic excitations
\cite{FL00}. More precisely, the number of atoms in the metastable
state $|m \rangle$ is exactly equal to that of the number of
incoming photons. Since the atomic excitation is collective, the
probability of finding any single atom in state $| m \rangle$
remains small. The state mapping process can in principle be
carried out with efficiency approaching $100 \%$. Initial
experiments already demonstrated reversible quantum state transfer
with efficiencies exceeding $20 \%$ \cite{hau01,walsworth}.

The principal idea of the proposal is that it is possible to
measure the number of atoms in state $| m \rangle$ very
accurately, using state-selective fluorescence technique used in
trapped ions \cite{ion-trap}. After turning the coupling laser
off, we turn on a "detection" laser field that is resonant with
the dipole allowed transition between state $\ m \rangle$ and
state $| f \rangle$. Just as in the ion trap experiments, we
choose states $|m \rangle$ and $| f \rangle$ to be $|F=2, m_F=2
\rangle$ and $|F=3, m_F=3 \rangle$ states of the ground (S) and
excited (P) levels respectively. The ground state $|g \rangle$
could then be the $|F=1, m_F=0 \rangle$ or the $|F=1, m_F=1
\rangle$ state, depending on the geometry of the coupling field.
Assuming $I=3/2$ nuclear spin typical for alkali atoms, this
choice guarantees that scattering of laser photons always project
the atoms back to state $| m \rangle$ and allows for multiple
photon scattering events.

To the extent that the number of atoms far exceeds the maximum
number of photons, the detection of scattered photons constitutes
a nearly ideal projective measurement of the photon number. To
highlight this fact, we can consider a (single atom) light
scattering time of $100 nsec$; for $\eta_s = 1 \%$ detection
efficiency of scattered photons, we expect to count $100$ photons
in $1 \mu sec$ for each atom in state $| m \rangle$. Assuming
Poisson statistics for scattered photons, we will then be able to
distinguish between photon number states $|n \rangle$ that have
small occupancy compared to $|n_{max} = 100 \rangle$. Naturally,
longer measurement times or higher $\eta_s$ will allow for higher
sensitivity and larger $n_{max}$. Since dephasing times of
hyperfine split states can be much longer longer than $1 msec$,
loss of dark-state coherence will not be important for a large
class of photon counting measurements.

The requirements for efficient quantum state transfer can be met
using either trapped cold or non-trapped warm atomic gases,
provided that Doppler free configuration is utilized in the
latter. The principal requirement is that the medium is optically
thick, i.e. $ \alpha = N \sigma / (2A) \gg 1$, where $\sigma$ is
the absorption cross-section of the $|g\rangle$-$|e\rangle$
transition (in the absence of the coupling field) and $A$ is the
cross-sectional area of the probe pulse. For a probe pulse
collimated to within $100 \mu m$ over $1 mm$ interaction region of
a cigar shaped cold atomic gas with $N = 10^5 $ atoms, this
condition is easily satisfied.

From a practical point-of-view, it would be a great simplification
to be able to use a warm atomic gas \cite{walsworth}. The
principal limitation of this set-up would be the off-resonant
absorption of the detection laser by the ground state atoms. The
maximum number of atoms allowed in the ground state will be
determined by the ratio of the off-resonant and resonant
scattering rates, for atoms in state $|g\rangle$ and $|m \rangle$,
respectively. Assuming that the ground-state hyperfine splitting
determines the detuning of the detection laser from state
$|g\rangle$ atoms, we find that $N_{max} \sim 10^6$ (in the
absence of Doppler broadening) . For $N \ge N_{max}$, the proposed
detector will effectively exhibit high dark counts.

Photon counting using dark-state polaritons in atomic media is
relatively slow due to the slow response time of atoms. The
bandwidth of the transparency window is given by $\Delta \nu =
\Omega_c^2(t) / (\Gamma_e \sqrt{\alpha})$, and determines the
shortest possible pulse-length of the probe field \cite{EIT}.
Here, $\Gamma_e$ is the spontaneous emission rate of the excited
state $|e \rangle$ (Fig.~1). State-selective fluorescence
technique requires that the atoms scatter many photons; having a
large $\Gamma_e$ will then speed up the photon counting process.
Given these considerations, one may conclude that implementation
of EIT and quantum-state transfer in semiconductors can lead to
faster photon counting \cite{EIT-semiconductor}.

As mentioned earlier, the state transfer method is reversible. If
the detection process had not altered the phase imprinted on the
atomic sample, it would have been possible to re-convert the
atomic collective excitation into probe photons. The overall
process would then have realized a quantum nondemolition (QND)
measurement of the photon number operator \cite{WM}, with nearly
unity efficiency. An interesting future direction would be to
determine if a photon-number QND measurement can be implemented
using the described techniques.

The fascinating proposal for implementing efficient quantum
computation using linear optics has two key requirements: (a)
deterministic generation of single-photon pulses, (b) photon
counting with efficiency exceeding $99 \%$ \cite{KLM01}. Since
significant advances in single-mode single-photon sources have
already been achieved \cite{michler}, an experimental realization
of the photon counter proposed in this Letter could enable the
implementation of basic elements of LOQC. One could argue that a
third requirement for LOQC is the possibility of storing
multi-mode multi-photon entangled states that are prepared
off-line; the quantum-state transfer technique that we utilize
would be an ideal candidate for that task. Finally, we note that
quantum state-transfer is a linear process; the nonlinearity
required for photon counting is provided by the anharmonicity of
the atomic spectrum, probed by the detection laser.

Photon counting in general, and single-photon detection in
particular has found many applications in basic physics
experiments. In particular, most of the Bell's inequality
violation experiments have utilized single-photon detection
\cite{aspect}. However, none of these experiments can eliminate
the so-called "detection loophole" \cite{aspect2} that arises from
low detection efficiency and could in principle allow for a local
realistic interpretation. The only Bell's inequality violation
experiment that avoided the detection loophole was in fact carried
out using trapped ions and the state-selective fluorescence
detection: this experiment however, cannot eliminate the
"lightcone loophole" \cite{ion-trap}. When realized, the photon
counter described in this Letter could allow for the first
loophole-free demonstration of Bell's inequality violation.

In summary, we have described a photon counter which realizes a
projective photon number measurement with efficiency approaching
$100 \%$. The device combines the experimentally demonstrated
techniques of state-selective fluorescence detection, and
quantum-state transfer between photons and collective atomic
excitations.

We thank S. E. Harris and S. Unlu for helpful discussions. This
research is supported by the Packard and Humboldt Foundations.

\begin{figure}[t]
\begin{center}
  \centerline{\psfig{figure=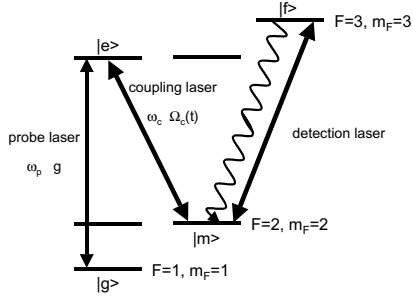,height=2.9in}}
\caption{The energy level diagram of the atoms used in quantum
state transfer and state-selective fluorescence detection. An
electromagnetically induced transparancy is established for the
probe field at frequency $\omega_p$ using the coupling field at
frequency $\omega_c$ and Rabi frequency $\Omega_c(t)$. When the
coupling field intensity is turned off adiabatically, each probe
photon will be stored as an atomic excitation in the metastable
state $|m \rangle$. By turning on a detection laser at the $|m
\rangle - |f \rangle$ transition, we can scatter many photons that
will reveal the number of atoms in state $|m\rangle$. The energy
levels are chosen assuming that the atomic gas is Doppler-free.}
\label{fig1}
\end{center}
\end{figure}

\end{document}